\documentclass{svproc}
\usepackage{graphicx}
\usepackage{amsmath,amssymb}
\def\hc{\text{h.c.}}
\usepackage{cite}

\usepackage{url}

\begin{document}
\mainmatter              
\title{Flavor and Dark Matter connection}
\titlerunning{Flavor and Dark Matter}  
%
\author{Avelino Vicente\inst{1}}
\authorrunning{Avelino Vicente} 
%
\tocauthor{Avelino Vicente}
\institute{Instituto de F\'{\i}sica Corpuscular (CSIC-Universitat de Val\`{e}ncia) \\
Apdo. 22085, E-46071 Valencia, Spain,\\
\email{avelino.vicente@ific.uv.es}}

\maketitle              

\begin{abstract}
In recent years, the LHCb collaboration has published results on the
measurement of several observables associated to semileptonic $b \to
s$ transitions. Interestingly, various deviations from their expected
values in the Standard Model have been found, including some
tantalizing hints pointing towards the violation of Lepton Flavor
Universality. We discuss New Physics models that address these
anomalies and explore their possible connection to the dark matter of
the Universe.
\end{abstract}
\section{Introduction}
\label{sec:intro}

One of the most relevant open questions in current physics is the
nature of the dark matter (DM) that mades up $27 \%$ of the energy
density of the Universe \cite{Ade:2015xua}. Many ideas and proposals
have been put forward, including the possibility that the DM is
composed of particles. This popular scenario requires an extension of
the Standard Model (SM) with new states and dynamics, since the SM
particles do not have the required properties to be a good DM
candidate.

Lepton Flavor Universality (LFU) is a central feature in the SM. The
fact that gauge bosons couple with the same strength to the three
generations of leptons is well rooted in the SM construction and has a
strong experimental support. Nevertheless, this expectation is broken
in some New Physics (NP) scenarios, and this can lead to clear
signatures of physics beyond the SM. In fact, these signatures might
have been observed already. Since 2013, the LHCb collaboration has
reported on the measurement of several observables associated to
semileptonic $b \to s$ transitions, finding some tensions with the SM
expectations, including possible hints of the violation of LFU.

We are interested in NP models that aim at an explanation of the
so-called {\it $b \to s$ anomalies} while introducing a dark sector
with the ingredients to accommodate the astrophysical and cosmological
indications of the existence of DM \cite{Vicente:2018xbv}. Several
authors have explored this direction
\cite{Sierra:2015fma,Belanger:2015nma,Allanach:2015gkd,Bauer:2015boy,Celis:2016ayl,Altmannshofer:2016jzy,Ko:2017quv,Ko:2017yrd,Cline:2017lvv,Sala:2017ihs,Ellis:2017nrp,Kawamura:2017ecz,Baek:2017sew,Cline:2017aed,Cline:2017qqu,Dhargyal:2017vcu,Chiang:2017zkh,Falkowski:2018dsl,Arcadi:2018tly,Baek:2018aru}. The
rest of the manuscript is organized as follows. We review the $b \to
s$ anomalies and provide a model independent interpretation in
Sec. \ref{sec:anomalies}. Sec. \ref{sec:classification} classifies the
proposed New Physics explanations to these anomalies with a link to
the dark matter problem and presnts two example models that illustrate
this connection. We finally summarize in Sec. \ref{sec:summary}.

\section{The $b \to s$ anomalies}
\label{sec:anomalies}

There are two types of $b \to s$ anomalies: (1) branching ratios and
angular observables, and (2) lepton flavor universality violating
(LFUV) anomalies.

\vspace*{0.2cm}

{\bf Branching ratios and angular observables:} In 2013, the LHCb
collaboration reported on the measurement of several observables in
the decay $B \to K^\ast \mu^+ \mu^-$ with $1$ fb$^{-1}$ of integrated
luminosity. Interestingly, several deviations with respect to the SM
expectations were found. The most popular one was a $3.7 \sigma$
discrepancy in one of the dimuon invariant mass bins in the
$P_5^\prime$ angular observable \cite{Aaij:2013qta}. Moreover, LHCb
also found a systematic deficit in several branching ratios, mainly
$\text{BR}(B_s \to \phi \mu^+ \mu^-)$ \cite{Aaij:2013aln}. These
anomalies were later confirmed with the addition of further data with
the presentation by LHCb of new results in 2015, using in this case
their full Run 1 dataset with $3$
fb$^{-1}$~\cite{Aaij:2015oid,Aaij:2015esa}.

\vspace*{0.2cm}

{\bf LFUV anomalies:} several observables have been proposed in order
to test LFU experimentally. In particular, one can consider the
$R_{K^{(\ast)}}$ ratios, given by~\cite{Hiller:2003js}
\begin{align}
R_{K^{(\ast)}} = \frac{ \Gamma(B \rightarrow K^{(\ast)} \mu^+ \mu^-)}{\Gamma(B \rightarrow K^{(\ast)} e^+ e^-)} \, .
\end{align}
These observables are measured in specific dilepton invariant mass
squared ranges $q^2 \in [q^2_{\rm min}, q^2_{\rm max}]$. These ratios
are very close to one in the SM, but this prediction can be altered by
NP violating LFU. Moreover, hadronic uncertainties are expected to
cancel to a high degree. Therefore, a deviation in $R_{K^{(\ast)}}$
would be regarded as a very clear sign of LFUV. Interestingly, several
measurements by the LHCb collaboration point in this direction. The
measurement of $R_K$ in the region $[1,6]$~GeV$^2$ was reported in
2014~\cite{Aaij:2014ora}, whereas the measurement of the $R_{K^\ast}$
ratio in two $q^2$ bins was announced in
2017~\cite{Aaij:2017vbb}. These were the results:
\begin{align}
R_K &= 0.745^{+0.090}_{-0.074}\pm0.036    \,, \quad
q^2 \in [1,6]~\text{GeV}^2 \,, \nonumber \\[0.2cm]
R_{K^\ast} &= 0.660^{+0.110}_{-0.070}\pm0.024    \,, \quad
q^2 \in [0.045,1.1]~\text{GeV}^2\,, \nonumber \\[0.2cm]
R_{K^\ast} &= 0.685^{+0.113}_{-0.069}\pm0.047    \,, \quad
q^2 \in [1.1,6.0]~\text{GeV}^2 \,. 
\end{align}
When these are compared to the SM predictions~\cite{Bordone:2016gaq},
\begin{align}
R_K^{\rm{SM}} &= 1.00 \pm 0.01   \,, \quad
q^2 \in [1,6]~\text{GeV}^2 \,, \nonumber \\[0.2cm]
R_{K^\ast}^{\rm{SM}} &= 0.92 \pm 0.02    \,, \quad
q^2 \in [0.045,1.1]~\text{GeV}^2 \,,\nonumber \\[0.2cm]
R_{K^\ast}^{\rm{SM}} &= 1.00 \pm 0.01    \,, \quad
q^2 \in [1.1,6.0]~\text{GeV}^2 \,,
\end{align}
one finds deviations from the SM at the $2.6\,\sigma$ level in the case of
$R_K$, $2.2\,\sigma$ for $R_{K^\ast}$ in the low-$q^2$ region, and
$2.4\,\sigma$ for $R_{K^\ast}$ in the central-$q^2$ region.

\vspace*{0.2cm}

While the first category, angular observables and branching ratios,
might be affected by hadronic uncertainties and the possibility of
uncontrolled QCD effects cannot be discarded, the second one, composed
by the $R_{K^{(\ast)}}$ ratios, is clean from this issue and can only
be explained by NP violating LFU.

In order to interpret these experimental results it is convenient to
adopt a language based on effective field theory. The effective
Hamiltonian for $b \to s$ transitions can be written as
\begin{equation} \label{eq:effH}
\mathcal H_{\text{eff}} = - \frac{4 G_F}{\sqrt{2}} \, V_{tb} V_{ts}^\ast \, \frac{e^2}{16 \pi^2} \, \sum_i \left(C_i \, \mathcal O_i + C^\prime_i \, \mathcal O^\prime_i \right) + \hc \, .
\end{equation}
Here $G_F$ is the Fermi constant, $e$ the electric charge and $V$ the
Cabibbo-Kobayashi-Maskawa matrix. The effective operators contributing
to $b \to s$ transitions are denoted by $\mathcal O_i$ and $\mathcal
O^\prime_i$, while $C_i$ and $C^\prime_i$ denote their Wilson
coefficients. The operators that turn out to be relevant for the
interpretation of the $b \to s$ anomalies are
\begin{align}
\mathcal O_9 &= \left( \bar s \gamma_\mu P_L b \right) \, \left( \bar \ell \gamma^\mu \ell \right) \, ,   &   \mathcal O^\prime_9 &= \left( \bar s \gamma_\mu P_R b \right) \, \left( \bar \ell \gamma^\mu \ell \right) \, , \label{eq:O9} \\
\mathcal O_{10} &= \left( \bar s \gamma_\mu P_L b \right) \, \left( \bar \ell \gamma^\mu \gamma_5 \ell \right) \, ,   &   \mathcal O^\prime_{10} &= \left( \bar s \gamma_\mu P_R b \right) \, \left( \bar \ell \gamma^\mu \gamma_5 \ell \right) \, . \label{eq:O10}
\end{align}
Here $\ell = e, \mu, \tau$. Unless necessary, we will omit flavor
indices in the Wilson coefficients in order to simplify the
notation. It proves convenient to split the Wilson coefficients into
their SM and NP pieces, defining
\begin{eqnarray}
C_9 &=& C_9^{\text{SM}} + C_9^{\text{NP}} \, , \\
C_{10} &=& C_{10}^{\text{SM}} + C_{10}^{\text{NP}} \, .
\end{eqnarray}

Several independent global fits
\cite{Capdevila:2017bsm,Altmannshofer:2017yso,DAmico:2017mtc,Hiller:2017bzc,Geng:2017svp,Ciuchini:2017mik,Alok:2017sui,Hurth:2017hxg}
have compared a large set of experimental measurements of observables
associated to $b \to s$ transitions to their expected values in the
SM, finding a remarkable tension, only alleviated by the introduction
of NP contributions. In particular, there is a general agreement on
the qualitative fact that global fits improve substantially with a
negative contribution in $C_9^{\mu , \text{NP}}$, with $C_9^{\mu ,
  \text{NP}} \sim - 25 \% \times C_9^{\mu , \text{SM}}$. NP
contributions in other muonic Wilson coefficients can affect the
global fit, but they are sub-dominant compared to $C_9^{\mu ,
  \text{NP}}$. For instance, the addition of NP in the one-dimensional
direction given by $C_9^{\mu , \text{NP}} = - C_{10}^{\mu ,
  \text{NP}}$ also serves to improve the fit, and this can be regarded
as a hint in favor of purely left-handed NP interactions. Moreover, no
hint for NP is found in contributions involving electrons or tau
leptons.~\footnote{See also \cite{Celis:2017doq} for a recent analysis
  of the $b \to s$ anomalies based on gauge invariant effective
  operators.}

\section{Model classification}
\label{sec:classification}

In general, the models explaining the $b \to s$ anomalies with a link
to the dark matter problem can be classified into two categories:

\begin{itemize}
\item {\bf Portal models:} in these models the mediator responsible
  for the NP contributions to $b \to s$ transitions is also the
  mediator for the production of DM in the early Universe.
\item {\bf Loop models:} in these models the required NP contributions
  to $b \to s$ transitions are induced via loops containing the DM
  particle.
\end{itemize}

There are also some {\it hybrid models} that share some properties
with both categories.

\subsection{A portal model}
\label{subsec:portal}

We will first discuss the portal model introduced in
\cite{Sierra:2015fma}. An extension of this model that also accounts
for neutrino masses has been recently discussed in
\cite{Rocha-Moran:2018jzu}.

{
\renewcommand{\arraystretch}{1.4}

\begin{table}
\centering
\begin{tabular}{ccccc} 
\hline \hline 
Field & Spin  & \( SU(3)_c \times\, SU(2)_L \times\, U(1)_Y \times U(1)_X \) \\ 
\hline
\(\phi\) & \(0\)  & \(({\bf 1}, {\bf 1}, 0, 2) \) \\ 
\(\chi\) & \(0\)  & \(({\bf 1}, {\bf 1}, 0, -1) \) \\ 
\(Q_{L,R}\) & \(\frac{1}{2}\)  & \(({\bf 3}, {\bf 2}, \frac{1}{6}, 2) \) \\ 
\(L_{L,R}\) & \(\frac{1}{2}\)  & \(({\bf 1}, {\bf 2}, -\frac{1}{2}, 2) \) \\ 
\hline \hline
\end{tabular} 
\caption{New scalars and fermions in the model of \cite{Sierra:2015fma}.}
\label{tab:DarkBS}
\end{table}
}


The model adds a new $U(1)_X$ factor to the SM gauge symmetry, with
its gauge boson denoted as $Z^\prime$ and its gauge coupling as
$g_X$. All the SM particles are singlets under this new symmetry,
while the new states beyond the SM, the vector-like (VL) fermions $Q$
and $L$ and the complex scalar fields $\phi$ and $\chi$, are
charged. Table \ref{tab:DarkBS} shows the charges of the new scalars
and fermions in the model. In addition to the usual canonical kinetic
terms, the new VL fermions $Q$ and $L$ have gauge-invariant
mass terms,
\begin{equation} \label{eq:VectorMass}
\mathcal L_m = m_Q \, \overline Q Q + m_L \, \overline L L \, .
\end{equation}
They also have Yukawa couplings with the SM doublets $q$ and $\ell$ and
the scalar $\phi$,
\begin{equation} \label{eq:VectorYukawa}
\mathcal L_Y = \lambda_Q \, \overline{Q_R} \, \phi \, q_L + \lambda_L \, \overline{L_R} \, \phi \, \ell_L + \hc \, .
\end{equation}
Here $\lambda_Q$ and $\lambda_L$ are $3$ component vectors. We will
consider that the scalar potential of the model leads to a vacuum
expectation value (VEV) for the $\phi$ scalar, $\langle \phi \rangle =
\frac{v_\phi}{\sqrt{2}}$, breaking $U(1)_X$ spontaneously and inducing
a mass for the $Z^\prime$ boson, $m_{Z^\prime} = 2 g_X v_\phi$. In
contrast, the scalar $\chi$ does not get a VEV. This leads to the
existence of a remnant $\mathbb{Z}_2$ parity, under which $\chi$ is
odd and all the other particles are even. This mechanism
\cite{Krauss:1988zc,Petersen:2009ip,Sierra:2014kua} stabilizes $\chi$
without the need of any additional symmetry.

\begin{figure}
\centering
\includegraphics[scale=0.4]{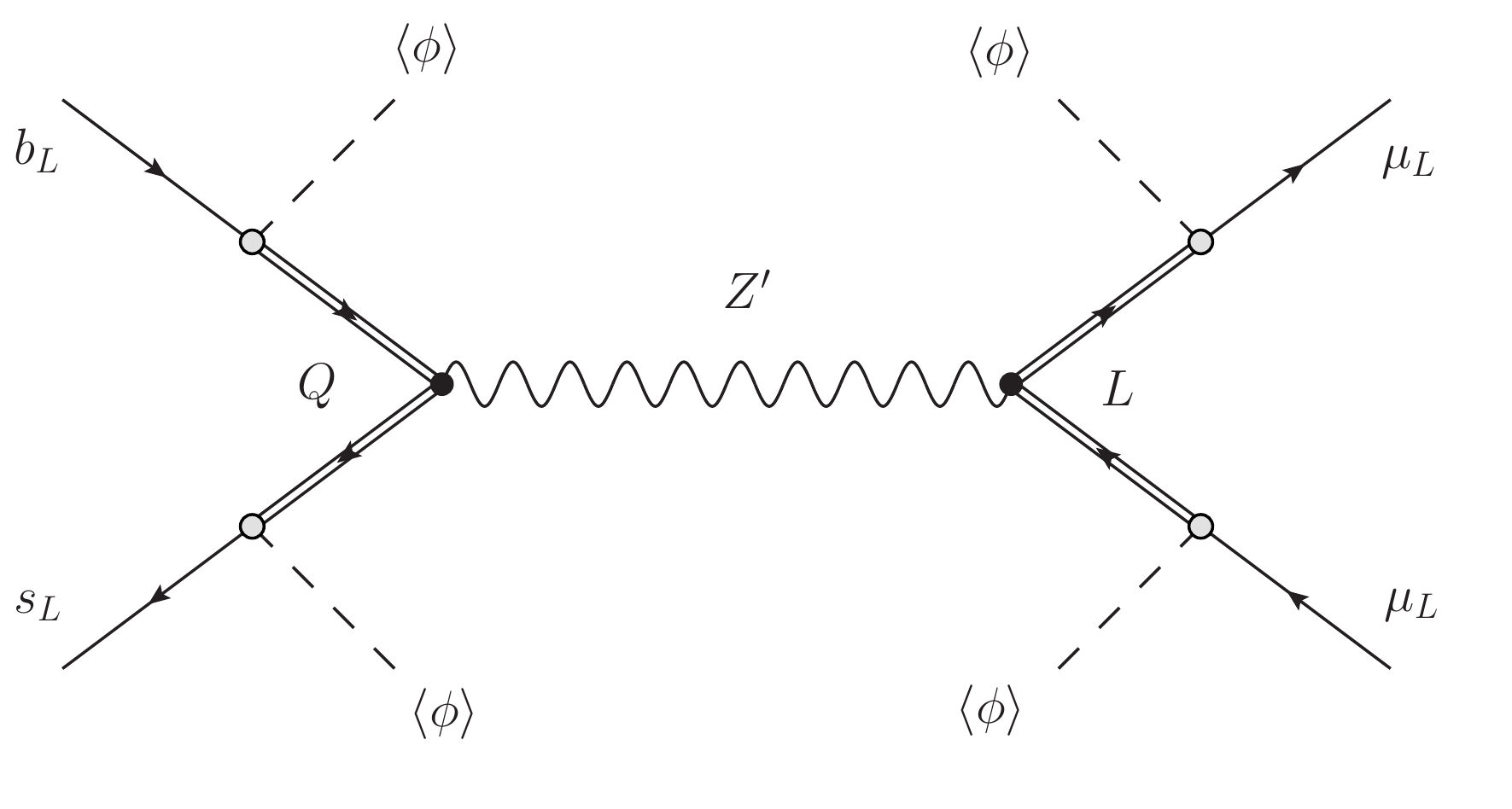}
\caption{Generation of $\mathcal O_9$ and $\mathcal O_{10}$ in the
  model of \cite{Sierra:2015fma}.}
\label{fig:couplings}
\end{figure}

The solution to the {\bf $\boldsymbol{b \to s}$ anomalies} in this
model is diagrammatically shown in Fig. \ref{fig:couplings}. The
Yukawa couplings in Eq. \eqref{eq:VectorYukawa} induce mixings between
the VL and SM fermions after $U(1)_X$ breaking. This mixing results in
$Z^\prime$ effective couplings to the SM fermions. Since the SM
fermions participating in the Yukawa interactions are purely
left-handed, the model predicts $C_9^{\mu , \text{NP}} = - C_{10}^{\mu
  , \text{NP}}$. It is possible to show that by using $|\lambda_L^\mu|
\sim 1 \gg |\lambda_Q^{b,s}|$ one can accommodate the required values
for $C_9^{\mu , \text{NP}}$ and $C_{10}^{\mu , \text{NP}}$ determined
by the global fits to $b \to s$ data and, at the same time, be
compatible with all constraints.

In what concerns to the {\bf Dark Matter} predictions of the model, we
already pointed out that $\chi$ is automatically stable due to the
remnant $\mathbb{Z}_2$ symmetry that is left after symmetry
breaking. Therefore, this is the DM candidate. Its production in the
early Universe takes place via $2\leftrightarrow 2$ processes mediated
by the massive $Z^\prime$ boson. This establishes a link with the $b
\to s$ anomalies and justifies the choice of name {\it portal models}
for the category represented by this model.

\subsection{A loop model}
\label{subsec:loop}


{
\renewcommand{\arraystretch}{1.4}
\begin{table}[t]
\centering
\begin{tabular}{cccccc} 
\hline \hline 
Field & Spin  & \( SU(3)_c \times\, SU(2)_L \times\, U(1)_Y \) & \( U(1)_X \) \\ 
\hline
\(X\) & \(0\)  & \(({\bf 1}, {\bf 1}, 0) \) & \(-1\) \\ 
\(Q_{L,R}\) & \(\frac{1}{2}\)  & \(({\bf 3}, {\bf 2}, \frac{1}{6}) \) & \(\phantom{+}1\) \\ 
\(L_{L,R}\) & \(\frac{1}{2}\)  & \(({\bf 1}, {\bf 2}, -\frac{1}{2}) \) & \(\phantom{+}1\) \\ 
\hline \hline
\end{tabular} 
\caption{New scalars and fermions in the model of \cite{Kawamura:2017ecz}.}
\label{tab:loopModel}
\end{table}
}

We now consider the model introduced in \cite{Kawamura:2017ecz}, a
simple illustration of the category of loop models. In this case, the
SM symmetry is extended with a global (not gauge) $U(1)_X$
symmetry. As in the previous model, all SM fields are assumed to be
singlets under this symmetry. In contrast, the new fields, the VL
fermions $Q$ and $L$ and the complex scalar $X$, are charged. Table
\ref{tab:loopModel} details the new scalar and fermionic fields and
their charges under the symmetries of the model.

The Lagrangian of the model contains the same Dirac mass terms as in
Eq. \eqref{eq:VectorMass}, as well as the Yukawa couplings
\begin{equation} \label{eq:VectorYukawa2}
\mathcal L_Y = \lambda_Q \, \overline{Q_R} \, X \, q_L + \lambda_L \, \overline{L_R} \, X \, \ell_L + \hc \, .
\end{equation}
Here $\lambda_Q$ and $\lambda_L$ are $3$ component vectors. We
consider a vacuum with $\langle X \rangle = 0$. This preserves the
global $U(1)_X$ symmetry and stabilizes the lightest state with a
non-vanishing charge under this symmetry. Furthermore, the
conservation of $U(1)_X$ forbids the mixing between SM and VL
fermions.

\begin{figure}
\centering
\includegraphics[scale=0.6]{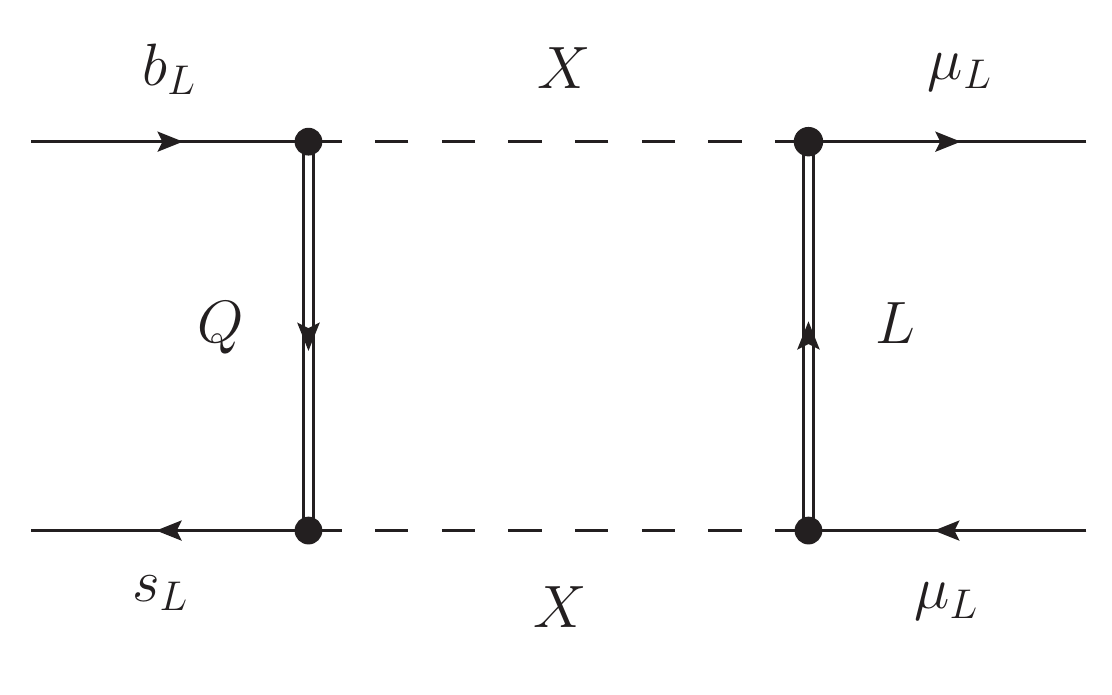}
\caption{Generation of $\mathcal O_9$ and $\mathcal O_{10}$ in the
  model of \cite{Kawamura:2017ecz}.}
\label{fig:loop}
\end{figure}

The solution of the {\bf $\boldsymbol{b \to s}$ anomalies} comes now
at the 1-loop level, as shown in Fig. \ref{fig:loop}. No NP
contributions to $b \to s$ transitions are generated at tree-level in
this model, as can be easily checked. As in the previous case, the
left-handed chirality of the fermions involved in the new Yukawa
interactions leads to $C_9^{\mu , \text{NP}} = - C_{10}^{\mu ,
  \text{NP}}$, and one can obtain the required ranges for these
Wilson coefficients by properly adjusting the parameters of the model.

Finally, we move on to the {\bf Dark Matter} phenomenology of the
model. Assuming that the lighest state charged under $U(1)_X$ is the
neutral scalar $X$, it constitutes the DM candidate in the model. As
discussed in detail in \cite{Kawamura:2017ecz}, the most relevant DM
annihilation channels for the determination of the DM relic density
are $X X^\ast \leftrightarrow \mu^+ \mu^-, \nu \nu$, and this is due
to the fact that one requires a large $|\lambda_L^\mu|$ in order to
account for the $b \to s$ anomalies. Interestingly, the model is also
testable in direct DM detection experiments, such as XENON1T.

\section{Summary}
\label{sec:summary}

Flavor and Dark Matter may seem two completely independent issues, but
they might be connected to the same fundamental physics. In these
proceedings we have discussed models that link the solution to the $b
\to s$ anomalies, a subject of great interest in current flavor
physics, to the existence of a dark sector. In doing this, these
models extend the SM with new ingredients, opening new model building
directions that would not be explored in the absence of this
connection. It would definitely be fascinating to find a deep bond
between these two areas of physics.

\section*{Acknowledgements}

I am very grateful to the organizers of the FPCP 2018 conference for
their hospitality and congratulate them for the exciting discussions
and friendly atmosphere during the meeting.  This work was supported
by the Spanish grants SEV-2014-0398 and FPA2017-85216-P (AEI/FEDER,
UE), PROMETEO/2018/165 and SEJI/2018/033 (Generalitat Valenciana) and
the Spanish Red Consolider MultiDark FPA2017‐90566‐REDC.

%
%

\end{document}